\documentstyle[11pt,aaspp4,flushrt]{article}

\slugcomment{\it Submitted to the Astrophysical Journal Letters}
\lefthead{Stern et al.}
\righthead{Mid-Infrared Selection of AGN}

\def\ie{{i.e.,~}}
\def\eg{{e.g.,~}}
\def\etal{{et al.}}

\def\deg{\ifmmode {^{\circ}}\else {$^\circ$}\fi}
\def\kms{\ifmmode {\rm\,km\,s^{-1}}\else
    ${\rm\,km\,s^{-1}}$\fi}
\def\ergcm2s{\ifmmode {\rm\,ergs\,cm^{-2}\,s^{-1}}\else
    ${\rm\,ergs\,cm^{-2}\,s^{-1}}$\fi}
\def\ergAcm2s{\ifmmode {\rm\,ergs\,cm^{-2}\,s^{-1}\,\AA^{-1}}\else
    ${\rm\,ergs\,cm^{-2}\,s^{-1}\,\AA^{-1}}$\fi}
\def\ergs{\ifmmode {\rm\,ergs\,s^{-1}}\else
    ${\rm\,ergs\,s^{-1}}$\fi}
\def\kmsMpc{\ifmmode {\rm\,km\,s^{-1}\,Mpc^{-1}}\else
    ${\rm\,km\,s^{-1}\,Mpc^{-1}}$\fi}

\def\lya{Ly$\alpha$}

\def\spose#1{\hbox to 0pt{#1\hss}}
\def\simlt{\mathrel{\spose{\lower 3pt\hbox{$\mathchar"218$}}
     \raise 2.0pt\hbox{$\mathchar"13C$}}}
\def\simgt{\mathrel{\spose{\lower 3pt\hbox{$\mathchar"218$}}
     \raise 2.0pt\hbox{$\mathchar"13E$}}}

\def\plotfiddle#1#2#3#4#5#6#7{\centering \leavevmode
\vbox to#2{\rule{0pt}{#2}}
\includegraphics{#1}}

\begin{document}


\title{Mid-Infrared Selection of Active Galaxies}

\author{Daniel Stern\altaffilmark{1},
Peter Eisenhardt\altaffilmark{1},
Varoujan Gorjian\altaffilmark{1}, 
Christopher S. Kochanek\altaffilmark{2},
Nelson Caldwell\altaffilmark{3},
Daniel Eisenstein\altaffilmark{4}, 
Mark Brodwin\altaffilmark{1},
Michael J.~I. Brown\altaffilmark{5,6},
Richard Cool\altaffilmark{4},
Arjun Dey\altaffilmark{6},
Paul Green\altaffilmark{3},
Buell T. Jannuzi\altaffilmark{6},
Stephen S. Murray\altaffilmark{3},
Michael A. Pahre\altaffilmark{3}, \&
S.~P. Willner\altaffilmark{3}} 

\altaffiltext{1}{Jet Propulsion Laboratory, California Institute of
Technology, 4800 Oak Grove Drive, Mail Stop 169-506, Pasadena, CA
91109 [e-mail: {\tt stern@zwolfkinder.jpl.nasa.gov}]}

\altaffiltext{2}{Department of Astronomy, The Ohio State University, 
Columbus, OH 43210}

\altaffiltext{3}{Steward Observatory, University of Arizona, 933 N.
Cherry Ave., Tucson, AZ 85721}

\altaffiltext{4}{Harvard-Smithsonian Center for Astrophysics, 60 Garden
Street, Cambridge, MA 02138}

\altaffiltext{5}{Princeton University Observatory, Princeton, NJ 08544}

\altaffiltext{6}{National Optical Astronomy Observatory, 950 N. Cherry
Avenue, Tucson, AZ 85719}

\begin{abstract} 

Mid-infrared photometry provides a robust technique for identifying
active galaxies.  While the ultraviolet to mid-infrared ($\lambda
\simlt 5 \mu$m) continuum of normal galaxies is dominated by the
composite stellar black body curve and peaks at approximately 1.6
$\mu$m, the ultraviolet to mid-infrared continuum of active galaxies is
dominated by a power law.  Consequently, with sufficient wavelength
baseline, one can easily distinguish AGN from stellar populations.
Mirroring the tendency of AGN to be {\em bluer} than galaxies in the
ultraviolet, where galaxies (and stars) sample the blue, rising portion
of stellar spectra, AGN tend to be {\em redder} than galaxies in the
mid-infrared, where galaxies sample the red, falling portion of the
stellar spectra.  We report on {\it Spitzer Space Telescope}
mid-infrared colors, derived from the IRAC Shallow Survey, of nearly
10,000 spectroscopically-identified sources from the AGN and Galaxy
Evolution Survey.  Based on this spectroscopic sample, we find that
simple mid-infrared color criteria provide remarkably robust separation
of active galaxies from normal galaxies and Galactic stars, with over
80\% completeness and less than 20\% contamination.  Considering only
broad-lined AGN, these mid-infrared color criteria identify over 90\%
of spectroscopically-identified quasars and Seyfert~1s.  Applying these
color criteria to the full imaging data set, we discuss the implied
surface density of AGN and find evidence for a large population of
optically-obscured active galaxies.

\end{abstract}


\section{Introduction}

The dominant sources of energy production in the universe are fusion in
stars and gravitational accretion onto supermassive black holes.  The
tight correlation between nuclear black hole mass and bulge mass
\markcite{Magorrian:98, Tremaine:02}(\eg Magorrian {et~al.} 1998; Tremaine {et~al.} 2002) implies the processes are
intimately connected.  However, identifying an unbiased census of black
holes in the universe remains challenging, hampering our ability to
fully probe this connection.  For example, the spectral index of the
X-ray background is significantly harder than the soft X-ray spectrum
of bright, unobscured quasars, implying that there exists a large
population of heavily-obscured AGN.  Identifying the sources
responsible for the (hard) X-ray background was one of the primary
motivators for the {\it Chandra X-ray Observatory}, and {\it Chandra}
has successfully identified a population of high-redshift,
heavily-obscured, luminous {\em type II quasars}
\markcite{Stern:02a, Norman:02}(\eg Stern {et~al.} 2002a; Norman {et~al.} 2002) as well as a significant population
of apparently-normal galaxies with optical spectra dominated by starlight but
hosting significant X-ray emission and implying the presence of an AGN
\markcite{Barger:01, Hornschemeier:01, Stern:02b}(\eg Barger {et~al.} 2001; Hornschemeier {et~al.} 2001; Stern {et~al.} 2002b).

We show here that the {\it Spitzer Space Telescope} \markcite{Werner:04}(Werner {et~al.} 2004)
also provides a valuable probe of AGN demographics.  The obscuring dust
which hides AGN from ultraviolet, optical, and soft X-ray surveys should
be a strong, largely isotropic emitter in the mid- to far-infrared
($\lambda \simgt 20 \mu$m).  Furthermore, the different spectral
energy distributions (SEDs) of stars and AGN produces different colors
for normal and active galaxies in the mid-infrared ($\lambda \simlt
10 \mu$m).  In particular, the composite black body spectra of the
stellar population of normal galaxies produces an SED which peaks at
approximately 1.6~$\mu$m, while quasars have a power-law SED, $f_\nu
\propto \nu^{-\alpha}$.  One of the early, traditional techniques to
identify quasars relied on identifying sources with an ultraviolet excess,
targeting sources whose SED does not plummet on the blue side of the
stellar peak \markcite{Schmidt:83}(\eg Schmidt \& Green 1983).  Though fruitful, this technique
fails in two cases: (1) obscured quasars will preferentially suffer
attenuation of their ultraviolet flux, and (2) high-redshift quasars
disappear from the ultraviolet due to absorption from the \lya\ forest.

By contrast, we seek to exploit the many-fold recent improvement in
mid-infrared sensitivity afforded by {\it Spitzer} to target sources
having SEDs that do not decline on the {\em red} side of the stellar
peak.  This provides an approach to identifying AGN that is relatively
insensitive to extinction by dust or gas, and it has been demonstrated
with a variety of instrumentation, including the Two Micron All-Sky
Survey \markcite{Cutri:01, Glikman:04}(Cutri {et~al.} 2001; Glikman {et~al.} 2004), the {\it Infrared Space
Observatory} \markcite{Laurent:00, Haas:04}(Laurent {et~al.} 2000; Haas {et~al.} 2004), and the First Look Survey
by {\it Spitzer} \markcite{Lacy:04}(Lacy {et~al.} 2004).

We present here a study of the mid-infrared colors of nearly 700
spectroscopically-confirmed quasars from the 9 deg$^2$ Bo\"otes field
of the NOAO Deep Wide-Field Survey \markcite{Jannuzi:99, Dey:05,
Jannuzi:05}(NDWFS; Jannuzi \& Dey 1999; Dey {et~al.} 2005; Jannuzi {et~al.} 2005).  The mid-infrared colors are from the IRAC Shallow Survey
\markcite{Eisenhardt:04}(Eisenhardt {et~al.} 2004), a guaranteed-time program with {\it Spitzer}.
The spectroscopy comes from the AGN and Galaxy Evolution Survey (AGES;
Kochanek \etal, in preparation).  AGES includes redshifts and spectral
classifications for nearly ten thousand normal galaxies and nearly one
thousand AGN.  Based on this spectroscopic sample, we show that
mid-infrared photometry provides a robust technique for identifying
AGN.  Using empirical color criteria to isolate AGN, we use the entire
IRAC Shallow Survey to study the surface density of active galaxies,
and evaluate the efficacy of mid-infrared selection of AGN.  A
coordinated paper discusses the mid-infrared colors of X-ray sources in
this field \markcite{Gorjian:05}(Gorjian {et~al.} 2005).

Section~2 briefly summarizes the two surveys used.  Section~3 describes
the mid-infrared properties of the spectroscopically-identified
sources, followed by a discussion of mid-infrared selection of AGN
in \S4.  The results are summarized in \S5.  Unless otherwise noted,
all magnitudes refer to the Vega system.

\section{Survey Data}

\subsection{IRAC Shallow Survey}

The Infrared Array Camera \markcite{Fazio:04a}(IRAC; Fazio {et~al.} 2004) is a four-channel
instrument on the {\it Spitzer Space Telescope} which provides
simultaneous broad-band images at 3.6, 4.5, 5.8 and 8.0~$\mu$m with
unprecedented sensitivity.  The IRAC shallow survey, a guaranteed-time
observation program of the IRAC instrument team, covers 8.5 deg$^2$ in
the NDWFS Bo\"otes field with three or more 30 second exposures per
position.  \markcite{Eisenhardt:04}Eisenhardt {et~al.} (2004) presents an overview of the survey
design, reduction, calibration, and initial results.  The survey
identifies $\approx$ 270,000, 200,000, 27,000 and 26,000 sources
brighter than $5\sigma$ limits of 12.3, 15.4, 76 and 76 $\mu$Jy at 3.6,
4.5, 5.8 and 8.0 $\mu$m, respectively, where throughout IRAC magnitudes
were measured in 6\arcsec\ diameter apertures and corrected to total
magnitudes assuming sources are unresolved at the $1\farcs66 -
1\farcs98$ resolution of IRAC \markcite{Fazio:04a}(Fazio {et~al.} 2004).  The corresponding
magnitude limits are 18.4, 17.7, 15.5, and 14.8 mag.

\subsection{AGN and Galaxy Evolution Survey (AGES)}

AGES (Kochanek \etal, in preparation) is a wide-field redshift survey
in the NDWFS Bo\"otes field using Hectospec \markcite{Fabricant:98}(Fabricant {et~al.} 1998), a
new, multi-object fiber spectrograph at the MMT Observatory.  For the
2004 observing season, AGES targeted sources from the NDWFS catalog
\markcite{Jannuzi:05}(Jannuzi {et~al.} 2005) with $R < 21.5$, with priorities based on the
multiwavelength photometry available.  AGES observed (1) all extended
sources with $R \leq 19.2$, (2) a randomly-selected 20\%\ of extended
sources with $19.2 < R \leq 20$, (3) all $R \leq 20$ extended sources with
IRAC 3.6, 4.5, 5.8 and 8.0 $\mu$m magnitudes $\leq$ 15.2, 15.2, 14.7 and
13.2, respectively, and (4) sources with a strong selection bias in favor
of counterparts to X-ray sources from {\it Chandra X-ray Observatory}
imaging of the field \markcite{Brand:05, Murray:05, Kenter:05}(Brand {et~al.} 2005; Murray {et~al.} 2005; Kenter {et~al.} 2005), radio
sources from the FIRST survey \markcite{Becker:95}(Becker, White, \& Helfand 1995), and 24~$\mu$m sources
with non-stellar $J -$ [24] colors, where the 24~$\mu$m data derives from
{\it Spitzer} Multiband Imaging Photometer \markcite{Rieke:04}(Rieke {et~al.} 2004) observations
of the field (Le~Floc'h \etal, in preparation).  Kochanek \etal\ (in
preparation) provides a more detailed description of the target selection.
In particular, point sources (in the $B_W$-, $R$-, or $I$-band images of
NDWFS) were selected against {\it unless} they were associated with $R
\leq 21.5$ counterparts to X-ray, radio, or 24~$\mu$m sources.

We use the version 1.11 AGES catalog, containing all AGES spectra taken
in 2004.  The redshift catalog, created by N.~Caldwell, P.~Green,
and C.~Kochanek, includes 10,452 sources with confident redshifts.
Non-stellar spectra were initially assigned to two rough classes:
broad-lined AGN and galaxies.  While all broad-lined sources correspond
to obvious AGN, \eg quasars and Seyfert~1s (Sy1s), identified galaxies
include both normal (\ie apparently inactive) galaxies and narrow-lined
AGN.  Fairly conservative emission line diagnostics\footnote{Objects fit
with the galaxy templates were flagged as AGN if they satisfied one of
the following criteria.  The first criterion was [NII]6583/H$\alpha >
0.5$, [OIII]5007/H$\beta > 2$, and detection of [OI]6300.   The second
criterion was the detection of [NeV]3426 emission.  The first criterion
can be used at lower redshifts ($z < 0.35$) while the second criterion can
be used at higher redshift ($z > 0.15$).} were then used to distinguish
AGN from normal galaxies.  The AGES sample with well-determined redshifts
is comprised of 9394 optically-inactive galaxies, 733 broad-lined AGN,
118 narrow-lined AGN, and 207 Galactic stars.  Most of the stars are F
stars, used for spectrophotometric calibration.  Matching these sources
with the 3.6~$\mu$m-selected IRAC Shallow Survey catalog by requiring
$\leq 2\arcsec$ separation between the IRAC mid-infrared source and NDWFS
optical source and $\geq 5\sigma$ detections in all four IRAC bands leads
to a restricted spectroscopic sample of 4693 sources, comprised of 3959
normal galaxies, 576 broad-lined AGN, 99 narrow-lined AGN, and 59 stars.
For some analyses, we only require IRAC 5$\sigma$ detections in the two
bluest IRAC bands; the matched catalog for this subsample contains 9432
sources, comprised (as classified by AGES) of 8460 normal galaxies,
696 broad-lined AGN, 104 narrow-lined AGN, and 172 stars.

\section{Mid-Infrared Properties of AGN}

\begin{figure}[!t]
\begin{center}
\plotfiddle{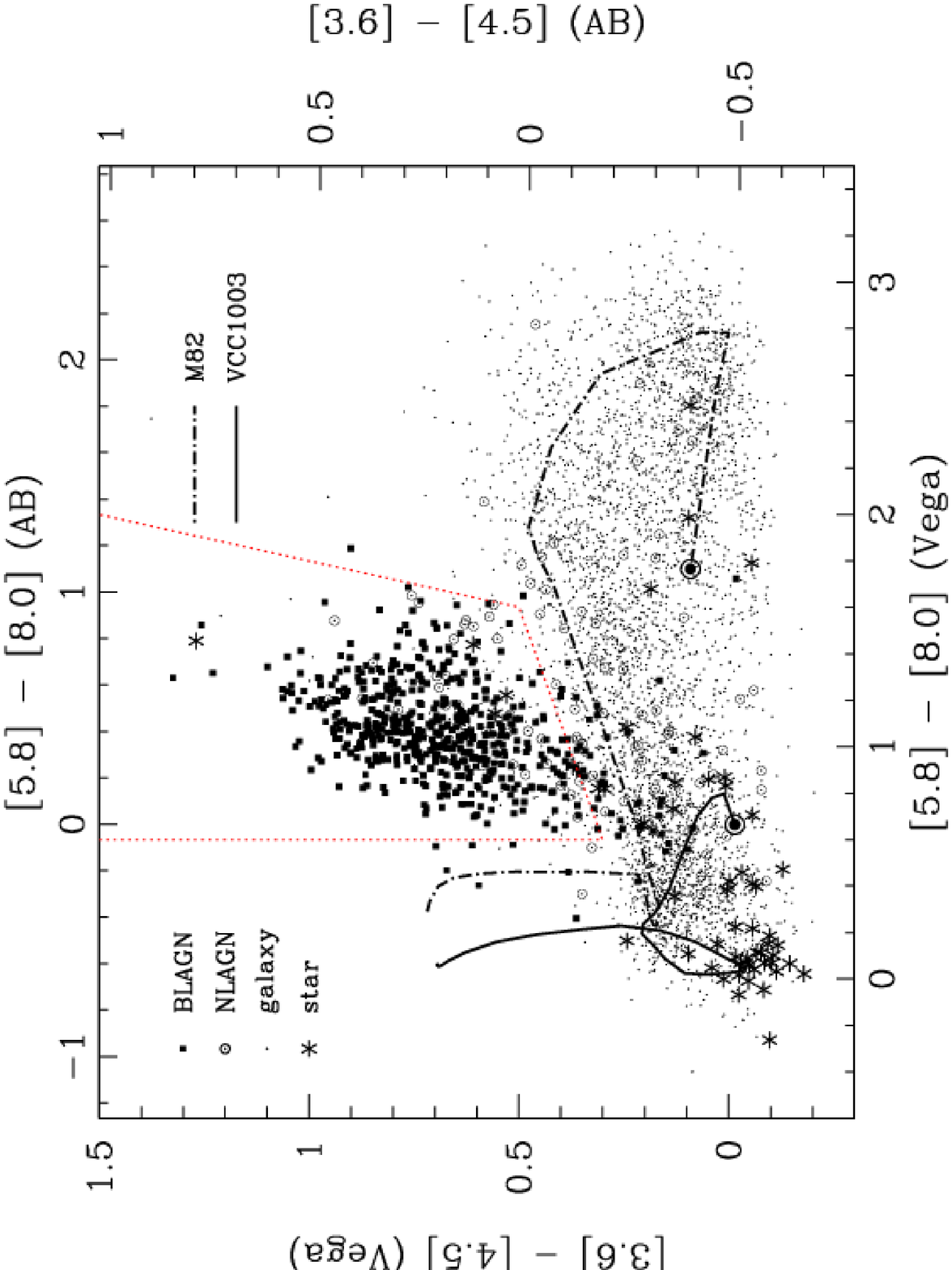}{6.2in}{-90}{65}{65}{-250}{390}
\end{center}

\caption{IRAC colors of spectroscopically-identified objects from the
AGES survey of the Bo\"otes field.  Axes indicate both the Vega and AB
magnitude systems.  Spectral classification of sources is annotated in
the upper left.  The $0 \leq z \leq 2$ color tracks for two
non-evolving galaxy templates from Devriendt \etal\ (1999) are
illustrated; dark bull's eyes indicate $z = 0$.  M82 is a starburst
galaxy, while VCC1003 (NGC~4429) is an S0/Sa galaxy with a star
formation rate approximately 4000 times lower.  The dotted line
empirically separates active galaxies from Galactic stars and normal
galaxies.}

\label{fig.colcol}
\end{figure}

Fig.~\ref{fig.colcol} presents an IRAC color-color diagram for the AGES
sample, with symbols denoting spectroscopic classification.
\markcite{Eisenhardt:04}Eisenhardt {et~al.} (2004), based solely on photometric data, noted a
vertical spur in the $[3.6] - [4.5]$ versus $[5.8] - [8.0]$ color-color
diagram (see Fig.~5c of that paper) and suggested that it may be
associated with AGN since a large fraction of those sources are
spatially unresolved at 3.6~$\mu$m.  These spectroscopic results
confirm that suggestion:  mid-infrared colors clearly separate AGN from
both stars and galaxies in the AGES sample.  In order to achieve a
physical understanding of Fig.~\ref{fig.colcol}, we now discuss the
mid-infrared properties of stars, galaxies, and AGN.

Galactic stars primarily reside in a restricted locus in mid-infrared
color space, corresponding to the Rayleigh-Jeans tail of the blackbody
spectrum:  stellar photospheres have approximately zero color in
the Vega system.  The majority of the spectroscopically-confirmed
stars in the AGES sample confirm this expectation.  The AGES sample
also includes several stars which reside far from the stellar locus,
showing mid-infrared excesses in the IRAC bands.  Such stars likely
host cooler circumstellar material in the form of post-asymptotic giant
branch shells or circumstellar disks.  A more detailed treatment of the
Galactic color-space outliers is deferred to a future publication.

\begin{figure}[!t]
\begin{center}
\plotfiddle{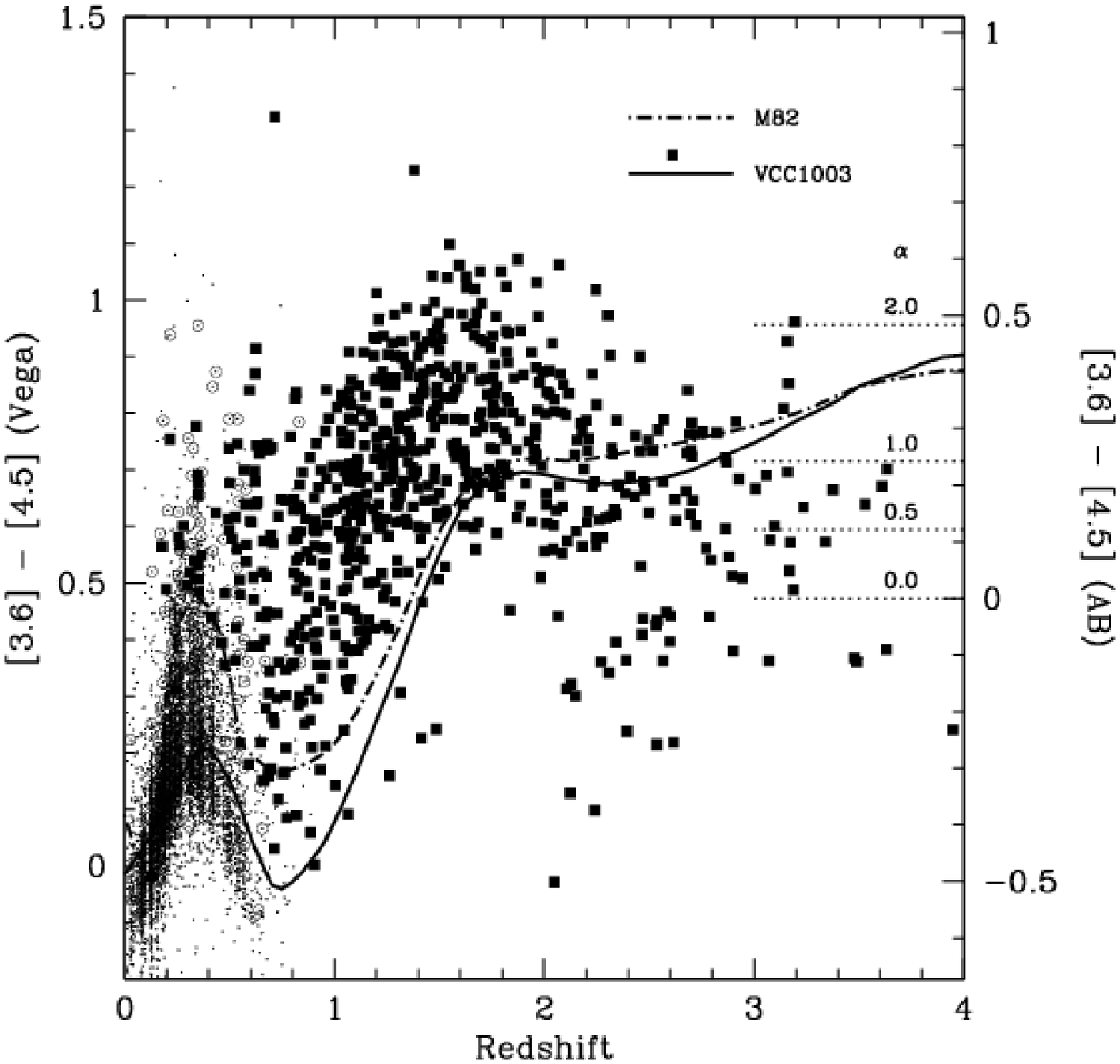}{2.5in}{0}{40}{40}{-240}{0}
\plotfiddle{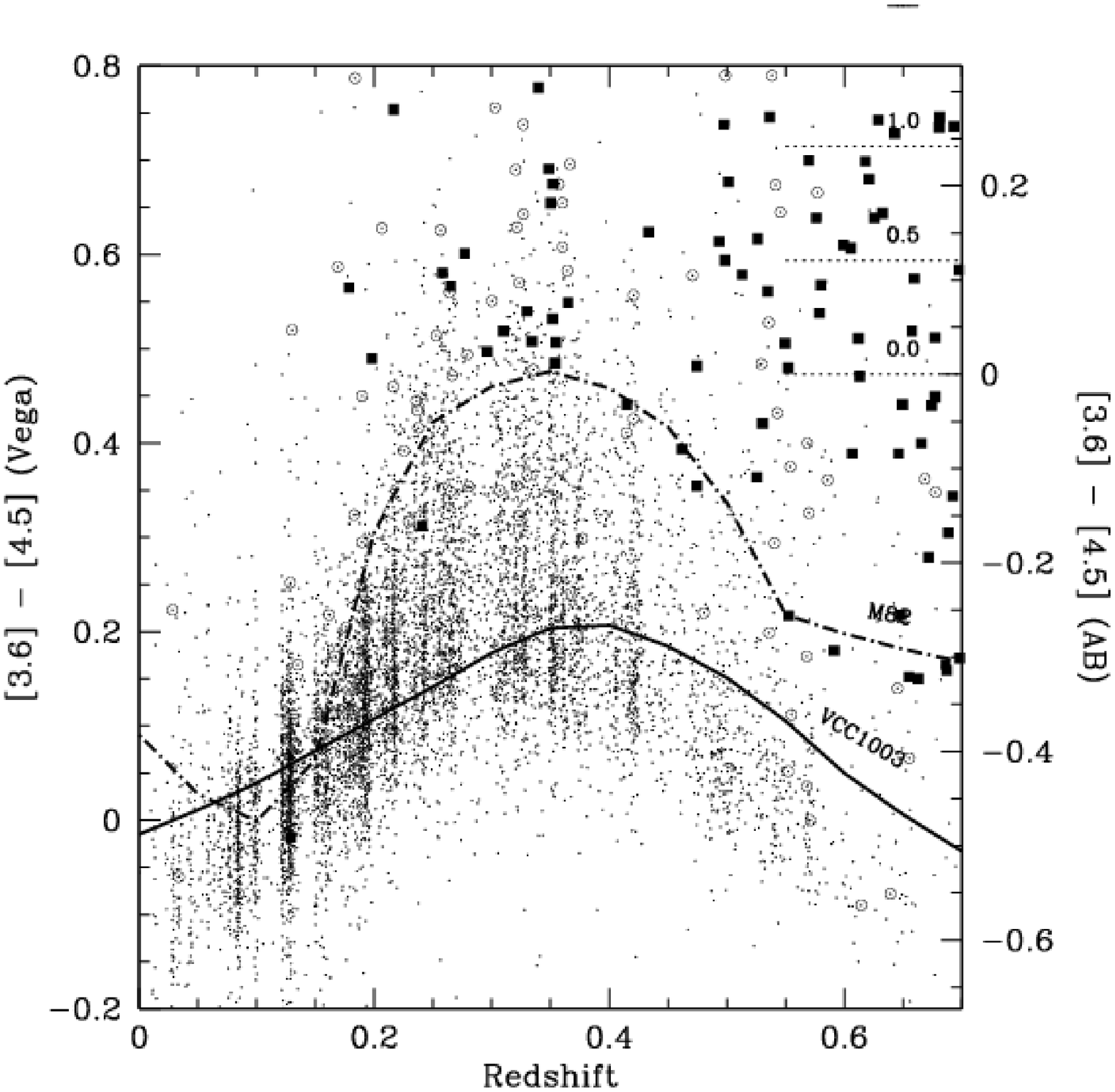}{0.0in}{0}{40}{40}{10}{30}
\end{center}

\caption{IRAC [3.6] $-$ [4.5] color evolution of galaxies, AGN, and
galaxy templates.  Symbols are the same as in Fig.~1.  The left-hand
panel shows the full observed redshift range, highlighting the modest
average color evolution of AGN.  The right-hand panel shows $z \leq
0.7$, highlighting the color evolution of normal galaxies.  Horizontal
dotted lines illustrate the expected mid-IR color of power-law SEDs,
$f_\nu \propto \nu^{-\alpha}$, for $0.5 < \alpha < 2.0$.}


\label{fig.ch12z}
\end{figure}

Because of its $R$-band magnitude limit, AGES identifies normal
galaxies out to $z \sim 0.6$.  As seen in Figs.~\ref{fig.colcol} and
\ref{fig.ch12z}, galaxies at $z \simlt 1$ span a relatively modest range
in the bluest IRAC color combination, $-0.2 \simlt {\rm [3.6]} - {\rm
[4.5]} \simlt 0.5$.  For reference, we show the color evolution of two
galaxy templates from \markcite{Devriendt:99}Devriendt, Guiderdoni, \&  Sadat (1999): M82 is a starburst galaxy
with strong polycyclic aromatic hydrocarbon (PAH) features, while VCC1003
(NGC~4429), a Virgo cluster S0/Sa galaxy, is not actively forming stars.
As the 3.3 $\mu$m PAH feature shifts through the IRAC 4.5 $\mu$m band at
$0.2 \simlt z \simlt 0.5$, galaxies appear redder in the ${\rm [3.6]} -
{\rm [4.5]}$ color combination.  At $z \simgt 1$, which lies beyond
the redshift range in which AGES is sensitive to normal galaxies, the
templates redden once again as the 1.6 $\mu$m maximum in the photospheric
emission from stellar populations shifts from the 3.6 $\mu$m IRAC band
to the 4.5 $\mu$m IRAC band.

\begin{figure}[!t]
\begin{center}
\plotfiddle{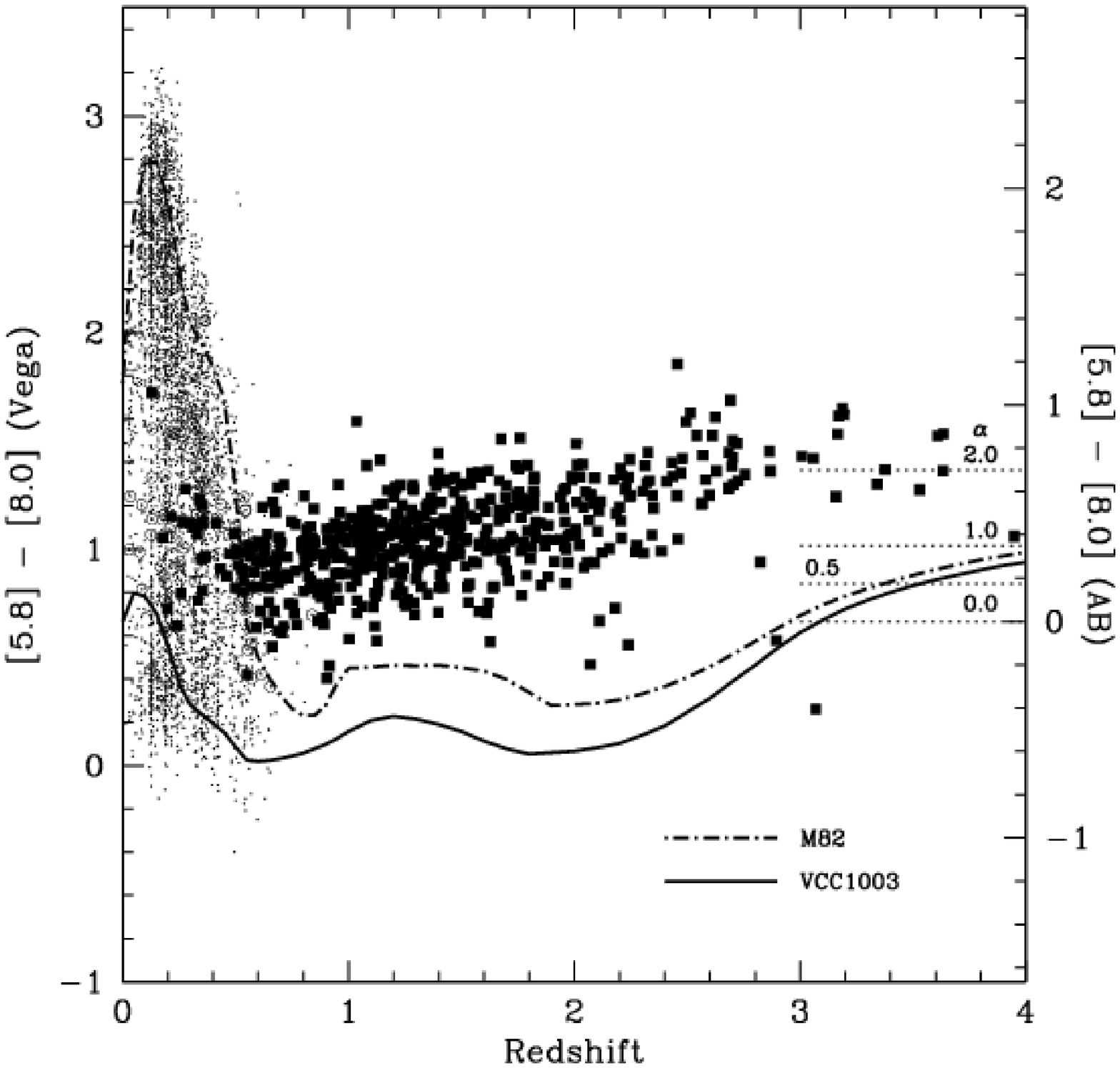}{2.5in}{0}{40}{40}{-240}{0}
\plotfiddle{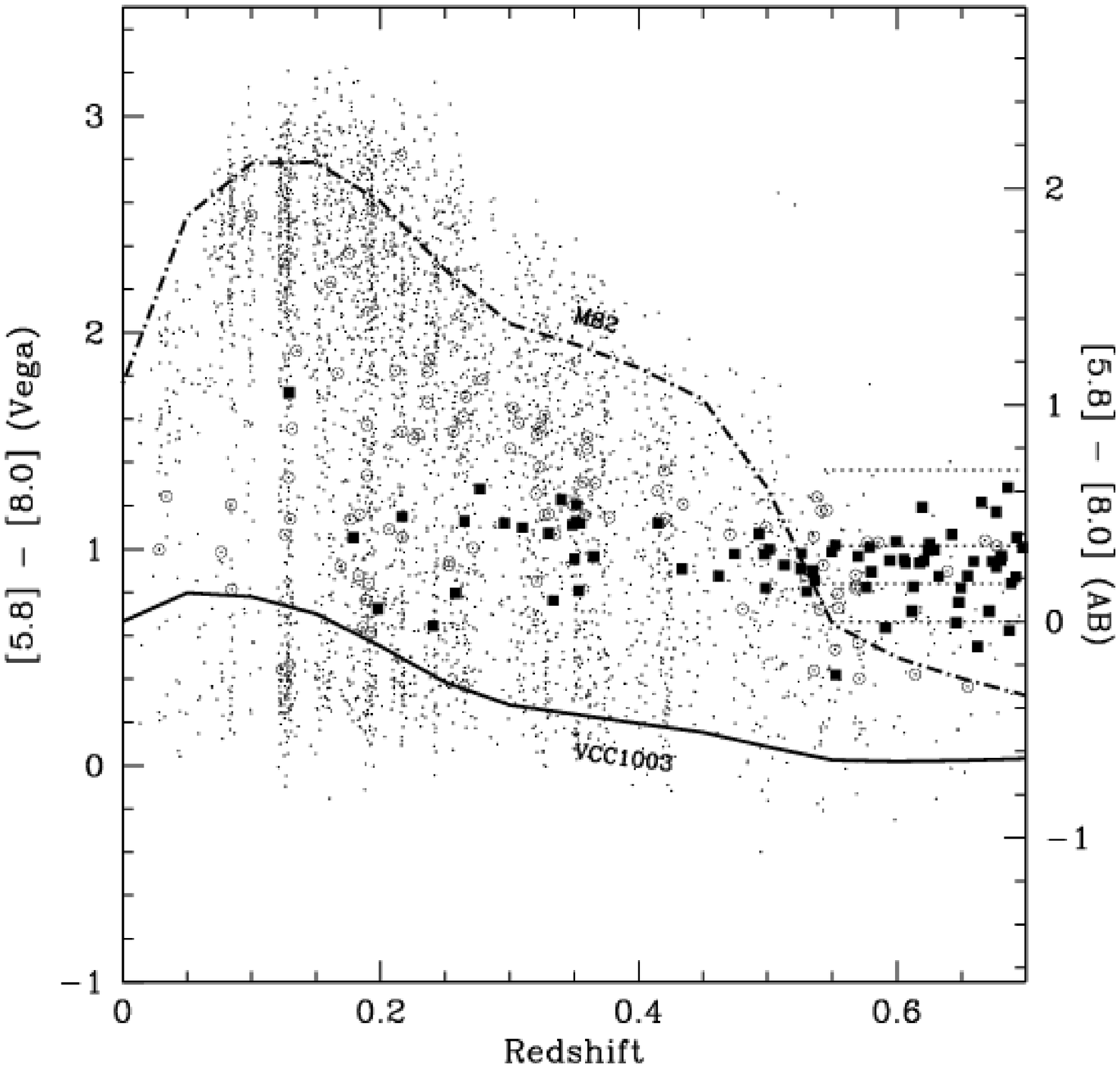}{0.0in}{0}{40}{40}{10}{30}
\end{center}

\caption{IRAC [5.8] $-$ [8.0] color evolution of galaxies, AGN, and
galaxy templates.  Symbols are the same as in Figs.~1 and 2.  The
left-hand panel shows the full observed redshift range, highlighting
the modest average color evolution of AGN.  The right-hand panel shows
$z \leq 0.7$, highlighting the color evolution of normal galaxies.
Horizontal dotted lines illustrate the expected mid-IR color of
power-law SEDs, $f_\nu \propto \nu^{-\alpha}$.}


\label{fig.ch34z}
\end{figure}

Galaxies at $z \simlt 0.6$ span a large range of color in the reddest
IRAC color combination, $0 \simlt {\rm [5.8]} - {\rm [8.0]} \simlt 3$
(Figs.~\ref{fig.colcol} and \ref{fig.ch34z}).  These extremely red
colors are caused by the 6.2 and 7.7 $\mu$m PAH features in actively
star-forming galaxies shifting through the IRAC 8.0 $\mu$m band.
The 6.2 and 7.7 $\mu$m PAH features are much stronger than the 3.3
$\mu$m PAH feature, explaining why the observed range of ${\rm [5.8]} -
{\rm [8.0]}$ colors for normal galaxies is much more expansive than the
observed ${\rm [3.6]} - {\rm [4.5]}$ color range.  The 3.3 $\mu$m PAH
feature later causes a modest, broad red bump in galaxy template ${\rm
[5.8]} - {\rm [8.0]}$ colors at $z \approx 1.3$.  At $z \simgt 2$, the
1.6 $\mu$m bump causes the galaxy templates to redden once again.  As seen
in Figs.~\ref{fig.colcol} thru \ref{fig.ch34z}, the \markcite{Devriendt:99}Devriendt {et~al.} (1999)
M82 and VCC1003 templates approximately bound the observed range of AGES
galaxies in both IRAC color$-$color space and in IRAC color$-$redshift
space.  The narrow-lined AGN in the AGES sample appear in both the
region of IRAC color$-$color space dominated by galaxies and 
the region dominated by broad-lined AGN, discussed next; the narrow-lined
AGN sample apparently includes sources whose mid-infrared flux can be
dominated by either stellar emission or emission associated with a
powerful active nucleus.

The isolation of broad-lined AGN in mid-infrared color-color space is
the most dramatic feature of Fig.~\ref{fig.colcol}.  While stars are
generally restricted to a single locus of zero color and $z \simlt 0.6$
galaxies define a horizontal swath in Fig.~\ref{fig.colcol},
broad-lined AGN create a vertical branch in the diagram.  This
characteristic property is easily understood.  First, the lack of
strong PAH emission in powerful AGN restricts their observed ${\rm
[5.8]} - {\rm [8.0]}$ colors.  Second, since the $\lambda \simlt
5~\mu$m flux of AGN is dominated by power-law emission rather than a
composite stellar spectrum which peaks at $\approx 1.6~\mu$m, the ${\rm
[3.6]} - {\rm [4.5]}$ color of AGN is significantly redder than that of
low-redshift galaxies.  The tendency of AGN to be {\em redder} than
galaxies in the mid-infrared, where galaxies sample the red, falling
side of the composite black-body stellar spectrum, mirrors the tendency
of AGN to be {\em bluer} than galaxies (and stars) in the ultraviolet,
where we are sampling the blue, rising side of the stellar spectrum.

The power-law nature of quasar SEDs is consistent with previous studies
at optical wavelengths.  \markcite{vandenBerk:01}{Vanden~Berk} {et~al.} (2001) presents a composite
quasar spectrum derived from over 2200 Sloan Digital Sky Survey (SDSS)
spectra.  Fitting the continuum at rest-frame wavelengths $\lambda
\lambda_0\, 1300 -5000$~\AA\ with a power law, $f_\nu \propto
\nu^{-\alpha}$, \markcite{vandenBerk:01}{Vanden~Berk} {et~al.} (2001) finds an average spectral slope
$\alpha = 0.44 \pm 0.1$, comparable to the $\alpha = 0.57 \pm 0.33$
derived by \markcite{Pentericci:03}Pentericci {et~al.} (2003) from a sample of 45 high-redshift SDSS
quasars imaged in the near-infrared.  IRAC observations of AGN from the
AGES survey suggest a modest steepening of the average quasar spectrum
at longer wavelengths.  Considering the 696 AGES broad-lined AGN with
robust ($\geq 5\sigma$) data in the bluest IRAC passbands, the
sigma-clipped average ${\rm [3.6]} - {\rm [4.5]}$ color is $0.65 \pm
0.20$ (Vega), corresponding to a spectra index $\alpha = 0.73 \pm
0.84$.  Considering the 576 AGES broad-lined AGN with robust photometry
in all four IRAC passbands, the sigma-clipped average ${\rm [5.8]} -
{\rm [8.0]}$ color is $1.06 \pm 0.21$ (Vega), corresponding to a
spectra index $\alpha = 1.13 \pm 0.59$.  We note that the reddening of
AGN at $z \simgt 1$ in observed ${\rm [3.6]} - {\rm [4.5]}$ color and
at $z \simgt 2$ in observed ${\rm [5.8]} - {\rm [8.0]}$ color is
suggestive of the rest-frame 1.6 $\mu$m stellar bump contributing to
the observed mid-infrared flux.  Kochanek \etal\ (in preparation)
presents a more detailed treatment of the composite AGES AGN spectrum
in the mid-infrared, derived from a principal component analysis of the
IRAC photometry.

\section{Mid-Infrared Selection of AGN}

We adopt the following empirical criteria to separate active galaxies
from other sources in the AGES spectroscopic sample described above
(shown as a dotted line in Fig.~\ref{fig.colcol}):
$$ \Bigl( {\rm [5.8]} - {\rm [8.0]} \Bigr) > 0.6 ~~~\wedge~~~
\Bigl( {\rm [3.6]} - {\rm [4.5]} \Bigr) >
0.2 \cdot \Bigl( {\rm [5.8]} - {\rm [8.0]} \Bigr) + 0.18$$
$$~~~\wedge~~~ \Bigl( {\rm [3.6]} - {\rm [4.5]} \Bigr) >
2.5 \cdot \Bigl( {\rm [5.8]} - {\rm [8.0]} \Bigr) - 3.5,$$
where $\wedge$ is the logical AND operator.  The left boundary protects
against fainter, higher-redshift galaxies, and is not critical for
optically-bright samples such as AGES.  The right boundary merely
approximates the outer range of AGN colors.  Considering Fig.~2,
we caution that these color criteria may preferentially omit AGN at
$z \approx 0.8$ and $z \approx 2$.  Out of a total of 681 sources
which reside in this portion of mid-infrared color-color space in
Fig.~\ref{fig.colcol}, 522 (77\%) are spectroscopically classified as
broad-lined AGN, 40 (6\%) are spectroscopically classified as narrow-lined
AGN, 113 (17\%) are spectroscopically classified as galaxies, and 6 (1\%)
are spectroscopically classified as stars.  Out of all 576 AGES sources
spectroscopically classified as broad-lined AGN and having good 4-band
IRAC data, 522 (91\%) meet these mid-infrared color criteria.  Of the 99
AGES sources spectroscopically classified as narrow-lined AGN and having
good 4-band IRAC data, 40 (40\%) meet these mid-infrared color criteria.
Furthermore, less than 3\% of the nearly 4000 AGES normal galaxies would
be (mis-)classified as AGN using these mid-infrared criteria.  We note
that sources classified as normal galaxies by AGES based on their optical
spectra may still host an active, supermassive black hole:  {\it Chandra}
has identified a large population of luminous X-ray sources ($L_{\rm
2-10\, keV} \simgt 10^{42}\, {\rm erg}\, {\rm s}^{-1}$) with apparently
normal optical spectra \markcite{Stern:02b}(\eg Stern {et~al.} 2002b).  Similar sources might be
below the existing X-ray limits for the Bo\"otes field \markcite{Brand:05,
Murray:05, Kenter:05}(Brand {et~al.} 2005; Murray {et~al.} 2005; Kenter {et~al.} 2005), but still be correctly identified as AGN by
{\it Spitzer}.

Optical surveys find an increasing number of quasars as a function of
flux --- the SDSS finds 15 quasars per deg$^2$ to $i^* = 19.1$
\markcite{Richards:02}(Richards {et~al.} 2002), the 2dF Quasar Redshift Survey finds 35 quasars
per deg$^2$ to $b_J = 20.85$ \markcite{Croom:04}(Croom {et~al.} 2004), and the COMBO-17 survey
AGN luminosity function implies approximately 90 quasars per deg$^2$ to
$R = 21$ \markcite{Wolf:03}(Wolf {et~al.} 2003).  Using the mid-infrared selection criteria on
the full IRAC Shallow Survey photometric catalog (ver.~1.1) of 14,099
non-saturated sources with $\geq 5\sigma$ detections in all IRAC bands,
there are 2014 AGN candidates over the 8.06 deg$^2$ of the NDWFS with
four-band IRAC coverage.  This translates to an approximate surface
density of 250 AGN candidates per deg$^2$ to a 8.0~$\mu$m flux limit of
76~$\mu$Jy.  For a spectral index $\alpha = 0.75$ (\S3), the
mid-infrared depth of the IRAC images correspond to $R \approx 21$.
The significantly higher surface density of IRAC-selected AGN relative
to optically-selected AGN of comparable mid-infrared flux implies a
sizable population of obscured quasars with attenuation of their
observed optical light.

The relatively shallow ($R \leq 21.5$) limit of the AGES spectroscopic
follow-up may bias our results:  AGES only identifies galaxies up to $z
\simlt 0.6$, while the \markcite{Devriendt:99}Devriendt {et~al.} (1999) SEDs plotted in
Fig.~\ref{fig.colcol} encroach upon our AGN selection criteria at $z
\approx 1.4$.  Some of the sources classified as AGN based on their
IRAC colors are likely moderate-redshift galaxies, fainter than the
AGES optical spectroscopic limits.  However, our implied AGN surface
density is below that predicted by models of mid-infrared AGN number
counts.  For $\Omega_\Lambda = 0.7$, $\Omega_m = 0.3$, and $H_0 = 70
\kmsMpc$, \markcite{Andreani:03}Andreani, Spinoglio, \& Malkan (2003) predicts $\sim 9 \times 10^4$ AGN per
deg$^2$ with an 8~$\mu$m flux higher than 100~$\mu$Jy, while
\markcite{Treister:04}Treister {et~al.} (2004) predicts only $\sim 1250$ AGN per deg$^2$ to
the same flux limit.  The large discrepancy is due to differences in
the assumed evolution of the AGN luminosity functions.  Since the IRAC
Shallow Survey only detects $\sim 3000$ {\em sources} per deg$^2$ with
an 8~$\mu$m flux higher than 76~$\mu$Jy, the \markcite{Andreani:03}Andreani {et~al.} (2003)
predictions are clearly high.

\section{Summary}

We analyze mid-infrared photometry from the IRAC Shallow Survey of 4693
sources with well-determined spectroscopic redshifts from version 1.11 of
AGES.  This sample includes 576 broad-lined AGN and 99 narrow-lined
AGN.  We find that mid-infrared photometry provides remarkably robust
separation of normal and active galaxies:  simple mid-infrared color
criteria identify over 90\% of the spectroscopically-identified
broad-lined AGN and 40\% of the spectroscopically-identified
narrow-lined AGN.  Our selection criteria rely on a mid-infrared excess
in AGN SEDs, mirroring the traditional ultraviolet excess technique
which has long been exploited to identify quasars.  Only 17\% of the
AGES sources which fall within our mid-infrared color criteria for AGN
are not spectroscopically classified as AGN.  Many of these sources
likely indeed host a luminous, active nucleus, where either dust
obscuration hides the signature of an AGN at observed optical
wavelengths, or the AGN signature is drowned out by the stellar
emission from the host galaxy \markcite{Moran:02}(\eg Moran, Filippenko, \& Chornock 2002).  Simple
mid-infrared selection criteria imply a surface density of 250 AGN
candidates per deg$^2$ to an 8~$\mu$m flux limit of 76~$\mu$Jy, nearly
triple the surface density inferred from optical surveys to this flux
limit.  AGN missing from optical surveys may be of the obscured, {\it
type II} variety which are implied by unified models of AGN and are
invoked by models of the X-ray background.  We show that the {\it
Spitzer Space Telescope} is a powerful tool for studying AGN
demographics; {\it Spitzer} will help identify a less-biased sample of
super massive black holes in the universe, allowing us to probe the
interconnection of fusion-driven and accretion-driven energy production
over cosmic history.

\acknowledgements 

We thank M.~Ashby and J.~Hora for carefully reading the manuscript, and
DS thanks E.~Treister for helpful discussion regarding models of AGN
number counts in the mid-infrared.  CSK thanks R.~Pogge for an
introduction to Seyfert galaxy identification.  This work is based on
observations made with the {\it Spitzer Space Telescope}, which is
operated by the Jet Propulsion Laboratory, California Institute of
Technology, under NASA contract 1407.  Support was provided by NASA
through an award issued by JPL/Caltech.  Spectroscopic observations
reported here were obtained at the MMT Observatory, a joint facility of
the Smithsonian Institution and the University of Arizona.  This work
also made use of images and/or data products provided by the NDWFS,
which is supported by the National Optical Astronomy Observatory
(NOAO).  NOAO is operated by AURA, Inc., under a cooperative agreement
with the National Science Foundation.

%
%


\clearpage
\end{document}